\documentclass[12pt]{iopart}

\usepackage{graphicx}

\begin{document}

\title[Ultra-bright and efficient single photon source]{Ultra-bright and efficient single photon generation based on N-V centres in
nanodiamonds on a solid immersion lens}

\author{Tim Schr{\"o}der, Friedemann G{\"a}deke, Moritz Julian Banholzer and Oliver Benson}

\address{Humboldt-Universit{\"a}t zu Berlin\\
Institut f{\"u}r Physik, AG Nano Optics\\
Newtonstr. 15, 12489 Berlin}

\ead{tim.schroeder@physik.hu-berlin.de}

\begin{abstract}
  Single photons are fundamental elements for quantum information
  technologies such as quantum cryptography, quantum information
  storage and optical quantum computing. Colour centres in diamond
  have proven to be stable single photon sources and thus essential
  components for reliable and integrated quantum information
  technology. A key requirement for such applications is a large
  photon flux and a high efficiency. Paying tribute to various
  attempts to maximise the single photon flux we show that collection
  efficiencies of photons from colour centres can be increased with a
  rather simple experimental setup. To do so we spin-coated
  nanodiamonds containing single nitrogen-vacancy colour centres on
  the flat surface of a ZrO$_2$ solid immersion lens. We found stable
  single photon count rates of up to 853\,kcts/s at saturation under
  continuous wave excitation while having excess to more than 100
  defect centres with count rates from 400\,kcts/s to 500\,kcts/s.
  For a blinking defect centre we found count rates up to 2.4\,Mcts/s
  for time intervals of several ten seconds. It seems to be a general
  feature that very high rates are accompanied by a blinking
  behaviour. The overall collection efficiency of our setup of up to 4.2\,\% is the highest yet reported for N-V defect centres in
  diamond. Under pulsed excitation of a stable emitter of 10\,MHz, 2.2\,\% of all
  pulses caused a click on the detector adding to 221\,kcts/s thus
  opening the way towards diamond based on-demand single photon
  sources for quantum applications.
\end{abstract}

\maketitle


Increased single photon collection with high on-demand efficiency
allows for higher communication bit rates in quantum cryptography
\cite{O'Brien2009} and faster read out of stationary qubits
\cite{Neumann2010,Dutt2007}. Also, a key operation in quantum
information processing (QIP), i.e. two-photon interference, requires
a large photon flux. Advanced protocols, such as entanglement
swapping or entanglement transfer \cite{Bouwmeester1997,Schmid2009}
also won't be possible if photons as flying qubits cannot be
generated or collected efficiently. A first crucial step concerns
reliable single photon generators. Such systems have been realised
based on atoms \cite{PhysRevLett.89.067901,Darquie2005}, ions
\cite{Keller2004}, molecules \cite{Lounis2000}, and solid state
based emitters like quantum dots \cite{Michler2000} or defect
centres in diamond \cite{PhysRevLett.85.290}. There were also
successful attempts to integrate solid-state emitters in photonic
nano-structures to increase the photon flux
\cite{springerlink:10.1140/epjd/e20020024,Claudon2010,Dousse2010}.
In particular defect centres in diamond have drawn a lot of
attention lately as they are photostable even at room temperature.
Since the first prove that single photons can be collected from
nitrogen-vacancy defect centres in diamond more than ten years ago
\cite{PhysRevLett.85.290}, steady improvement of collection rates
has been achieved. Furthermore other defect centres with different
optical behaviour have been found and fabricated
\cite{simpson:203107,Wang2006,rabeau:134104} in bulk as well as in
diamond nanocrystals. Nanodiamonds have several advantages compared
to bulk diamond as they can be easily deposited on various
substrates with e.g. spin-coating techniques or nano-manipulation.
AFM manipulation has been used to integrate them on photonic crystal
resonators \cite{wolters:141108} or on optical fibres
\cite{schroeder2010}. Moreover photon collection efficiency of
nanodiamonds is generally higher \cite{Beveratos2002}. Yet it has
just recently been shown, that reduction of excitation intensity and
improvement of emission collection of N-V centres in bulk diamond
can be achieved via etching adequate structures into CVD grown bulk
diamond \cite{Babinec2010,Hadden2010,Siyushev2010}. These structures
can be nanorods \cite{Babinec2010} or solid immersion lenses built
into bulk diamond \cite{Hadden2010} or made out of bulk diamond
\cite{Siyushev2010} and facilitate collection of single photons of
up to 500\,kcts/s. Unfortunately production of these structures
demands rather sophisticated processes.

Solid immersion lenses (SIL) on the other hand have been integrated
into optical microscopes 1990 \cite{Mansfield1990} and are
commercially available in various designs. A solid immersion lens
can be formed as a simple half-sphere or in a so called
Weierstrass-design \cite{Terris1994} and is most typically made of
high index material, e.g. ZrO$_2$ with $n=2.17$. The operation
principle of a SIL is very similar to an oil-immersion microscope.
Samples are either deposited on or close to the flat side of the SIL
while the optical access lies on the curved side of the SIL. In this
manner the wave front of a focused light beam passes the surface
parallel and neither refraction nor aberration is taking place. Due
to the high index of refraction the numerical aperture is enhanced
warranting higher collection angles and thus higher emitter
collection rates. Also the resolution is increased similar as in
oil-immersion-microscopes and much higher energy densities in the
focus are obtained \cite{Mansfield1990}. With respect to the
implementation of SILs with a single photon emitter there is another
important feature. On account of the strong step in the index of
refraction at the SIL-air interface the dipole emission is not
symmetrically distributed into SIL and air respectively
\cite{koyama:1667}. We performed finite difference time domain
(FDTD) calculations for our specific system and found that a simple
dipole emitter in air, 10 nm away from a flat material surface with
$n=2.17$, emits more than 86\,\% of its emission into the direction
of the SIL if oriented parallel to the surface and more than 75\,\%
if oriented perpendicular to the surface. These results are in
agreement with an analytical analysis of a dipole emitter in front
of a flat dielectric interface \cite{Lukosz:77}. The combination of
commercial availability, easy experimental integration, well
increased NA, smaller focus and strongly enhanced emission into
direction of collection, make a solid immersion lens a very potent
tool to increase single photon emitter collection efficiencies.

\begin{figure}
  \centering
  \includegraphics[width=0.667\textwidth]{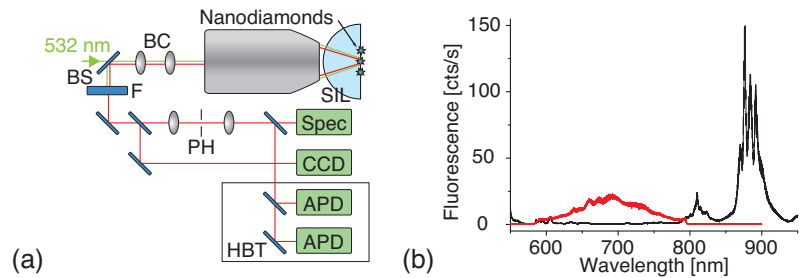}
  \caption{Experimental setup and spectroscopic properties of the solid immersion lens.
  (a) Sketch of home-built confocal microscope. BC, BS, F, PH, HBT stand for beam control, beam splitter, filter, pinhole, Hanbury Brown
  and Twiss setup, respectively.
  (b) Spectrum of the solid immersion lens fluorescence (black curve) at excitation power of 933\,$\mu$W and spectrum of N-V centre
  on SIL, emission cut off by 590\,nm long pass and 795\,nm short pass filters (red curve).}
  \label{fig:1}
\end{figure}

Here we present the experimental implementation of a SIL microscope
combined with the advantages of nanodiamonds which outperform so-far
reported single photon count rates from N-V centres in diamond. By
spin-coating nanodiamonds on the flat surface of commercially
available ZrO$_2$ half-spheres we prepared a SIL suitable to
efficiently excite colour centres inside the nanodiamonds. The SIL
was implemented into a home-built confocal microscope featuring an
air objetive with a numerical aperture of 0.9 and a pinhole of
100\,$\mu$m as depicted in figure \ref{fig:1}(a). To reduce stray
light from laser excitation and SIL fluorescence 540\,nm and 590\,nm
long pass filters as well as a 795\,nm short pass filter were used.
Via the spin-coating technique there is a wide flexibility to
integrate nanodiamonds with arbitrary colour centres into any kind
of SIL. The latter can be half-sphere SILs or Weierstrass SILs made
of various materials transparent at excitation and colour centre
emission wavelength. The chosen material ZrO$_2$ has a high index of
refraction $n$ of 2.17 at 600\,nm and can be processed by diverse
companies to half-spheres within required tolerances
\cite{baba:6923}. Its intrinsic fluorescence due to impurities is
sufficiently negligible between wavelengths of 615\,nm and 785\,nm
at typical excitation powers for defect centres in diamond of less
than 1\,mW (see spectrum in figure \ref{fig:1}(b)). It is well
suitable for the collection of photons from e.g. N-V centres
\cite{PhysRevLett.85.290}, Ni/Si related centres
\cite{PhysRevB.79.235316,Steinmetz2010} and Si-V centres
\cite{Wang2006,Neu2010} emitting from 600\,nm to 800\,nm (see
spectrum of a N-V centre in figure \ref{fig:1}(b)), around 770\,nm
and around 740\,nm, respectively.

\begin{figure}
  \centering
  \includegraphics[width=0.667\textwidth]{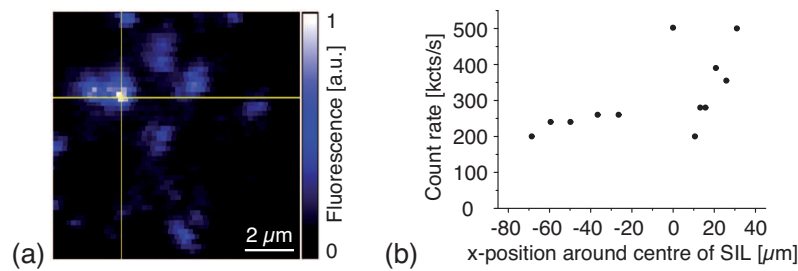}
  \caption{x-y-confocal scan of nanodiamond emission on SIL and distribution of N-V centre intensity. (a) 10\,$\mu$m by 10\,$\mu$m
  intensity scan of an area close to the centre of the SIL. Such area scans most likely feature a couple bright
  ($R_\mathrm{Inf} = $400\,kcts/s to 700\,kcts/s) and many darker N-V centres ($R_\mathrm{Inf}<$400\,kcts/s).
  (b) Count rate of different N-V centres under pulsed excitation of 80\,MHz and excitation intensities of 148\,$\mu$W in the
  focus. 12 N-V centres along one scan direction (x-axis) were
  investigated. All emitters provide single emitter characteristics proven by a measured value of the normalized $2^\mathrm{nd}$ order correlation function
  $g^{(2)}(0)<0.5$.}
  \label{fig:2}
\end{figure}

\begin{figure}
  \centering
  \includegraphics[width=1.0\textwidth]{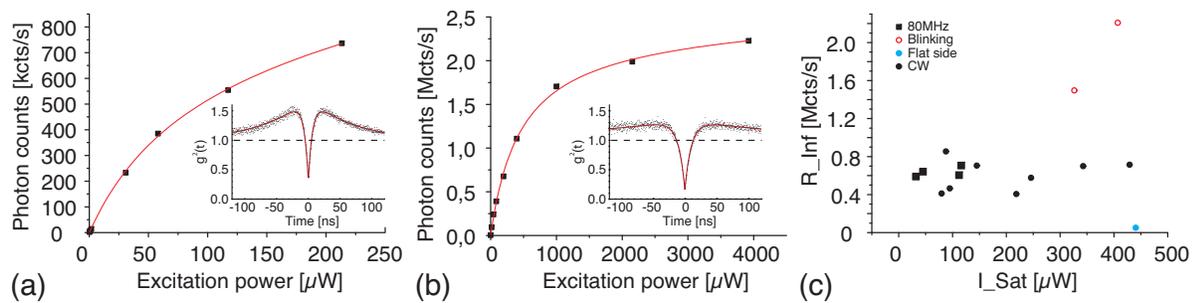}
  \caption{Autocorrelation and saturation measurements of different N-V centres.
  (a) Saturation measurement of the brightest stable emitter  found. $R_\mathrm{Inf}$ is 853\,kcts/s while saturation
  excitation intensity is 88\,$\mu$W. The inset shows the normalised autocorrelation function $g^{(2)}(\tau)$ of the same
  emitter with $g^{(2)}(0)<0.3$. (b) Saturation measurement of the brightest blinking emitter  found. $R_\mathrm{Inf}$ is
  2.4\,Mcts/s on a time basis of several ten seconds while saturation excitation intensity is 464\,$\mu$W. The averaged count rate is
  477\,kcts/s for an excitation intensity of 2.6\,mW. The inset shows the normalised autocorrelation function $g^{(2)}(\tau)$
  of the same emitter with $g^{(2)}(0)<0.16$ at excitation of 1\,mW. The red curves in (a) and (b) are theoretical fits to the data (see text).
  (c) Saturation count rates of different N-V centres as function of saturation excitation intensities. Black symbols represent stable emitters,
  open red circles blinking ones and the blue dot represents a N-V centre probed from the flat side of the SIL (the right side of the SIL in figure \ref{fig:1}(a)).
  Square symbols denote quasi-CW excitation with 80\,MHz laser repetition rate while circles relate to CW excitation.}
  \label{fig:3}
\end{figure}

For the applied spin-coating process an aqueous solution with
0.01\,\% polyvinyl alcohol and nanodiamonds with a mean size of
25\,nm was prepared. Spin-coating of a solution with suitable
diamond density was performed with 2000\,rpm to ensure a dense
distribution of the diamonds with distances smaller 1\,$\mu$m.
Figure \ref{fig:2}(a) shows a x-y-intensity scan of the confocal
microscope of a 10\,$\mu$m x 10\,$\mu$m region with typical diamond
distribution and fluorescence intensity. We found that in a
rectangular array of about 10\,$\mu$m by 100\,$\mu$m around the
centre of the SIL at least 10 diamonds have count rates of more than
250\,kcts/s if excited with 148\,$\mu$W of quasi-CW excitation with
laser repetition rate of 80\,MHz while many more had count rates
around 200\,kcts/s (see figure \ref{fig:2}(b)). It should be
mentioned that many of the diamonds investigated showed blinking
behaviour. Comparing these count rates to saturation measurements
(see e.g. figure \ref{fig:3}(a)) we extrapolate count rates at
saturation under CW excitation of 400\,kcts/s to 500\,kcts/s for the
10 measured defect centres. For an area of 100\,$\mu$m by
100\,$\mu$m we expect more than 100 colour centres with similar
emission rates. In this manner suitable diamonds for special
applications can be selected, e.g. one diamond with maximum
collection rates or two diamonds with matching zero phonon line
emission for two-photon interference experiments.

To prove single photon character of studied colour centre emissions
we performed autocorrelation measurements in a Hanbury Brown and
Twiss (HBT) setup (see figure \ref{fig:1}(a)). If the normalized
intensity autocorrelation function
$$g^{(2)}(\tau) = \frac{ \langle I(t)\,I(t+\tau) \rangle }{ \langle I(t) \rangle^2 }$$
has values at zero time delay $\tau = 0$ of $g^{(2)}(0)<0.5$,
emission should occur basically from a single N-V centre. Typical
$g^{(2)}(\tau)$ functions of the measured N-V centres have a
$g^{(2)}(0)$ between 0.1 and 0.3, depending on excitation
intensities. In order to determine the maximum accessible photon
flux we performed saturation measurements (displayed in figure
\ref{fig:3}(a),(b)). Fits to the experimental curves where done
according to
$$R (I) = \frac{ R_\mathrm{Inf}\ I }{ I_\mathrm{Sat}+I }+(A+\alpha)I+\beta$$

where $R$ is the single photon count rate, $R_\mathrm{Inf}$ the
count rate at infinite excitation intensities, $I$ the excitation
intensity, $I_\mathrm{Sat}$ the saturation excitation intensity, $A$
represents the measured background fluorescence 1\,$\mu$m away from
the N-V centre, while $\alpha$ and $\beta$ are fit parameters for
linear background stemming from the diamond and additional
background such as APD dark counts and residual stray light,
respectively. Most saturation excitation intensities were
distributed around 80\,$\mu$W in the focus while the typical single
photon count rate $R$ was 500\,kcts/s without background
substraction as can be seen in figure 3(c). Twelve single N-V
centres of similar brightness were analyzed. It should be pointed
out though that these N-V centres have quite different saturation
intensities as can be derived from figure 3(c). This is to some
degree due to their randomly distributed dipole orientations.
Furthermore the nanodiamonds have mismatched physical and optical
contact to the SIL surface resulting in diverse optical coupling of
the nanodiamonds to the SIL. In figure 3(c) also the count rate at
saturation for a random N-V centre that has been probed from the air
side of the SIL is depicted. This N-V centre, the brightest we found
from the flat side, shows count rates at saturation of 50\,kcts/s,
about ten times lower intensity than those excited through the SIL.
Furthermore its saturation intensity of 440\,$\mu$W is also about 5
times higher compared to these N-V centres.

We found one outstanding N-V centre with a stable single photon
count rate at saturation of 853\,kcts/s. Even at excitation
intensities of 213\,$\mu$W and count rates of 736\,kcts/s its
$g^{(2)}(0)$ value was smaller than 0.3 and could be further reduced
by adequate filtering (see figure \ref{fig:1}(b)). Furthermore we
observed another remarkable single defect centre. This defect centre
was ultra bright as we found single photon rates up to 2.4 Mcts/s at
saturation on a time base up to several ten seconds alternating with
darker periods. On the average, the single photon emission rate was
400\,kcts/s at excitation power of 2.6\,mW. We also measured its
autocorrelation function at count rates of up to 1.1\,Mcts/s to be
smaller than $g^{(2)}(0)<0.16$. In our experiments we found that in
an ensemble of nanodiamonds there exist a few ultra-bright N-V
defect centres with rates $> 1$\,Mcts/s. However, this large rate
seems to be accompanied by pronounced blinking behaviour. The reason
for this is not yet understood.

\vspace{\baselineskip}

For most applications of single photons in QIP an on-demand
generation is required \cite{Ladd2010}. One possibility is to use
pulsed excitation. A crucial parameter of such a source is the
collection efficiency per excitation pulse. A perfect single photon
on-demand device would deliver exactly one detected single photon
per pulse in a well defined optical mode. Such performance is
limited in real devices mainly by the efficiency to collect single
photon emission with the first lens, losses in the optical pass
towards the detector as well as by detector efficiencies.

To determine the collection efficiency of our setup we measured the
collection efficiency per excitation pulse. The repetition rate of
the pulsed excitation laser was limited to about 10\,MHz, such that
the time interval between pulses was 100\,ns, i.e. much larger than
the lifetime of 18\,ns of the N-V centre. With these numbers the
probability for excited N-V centres to decay before the arrival of
the next exciting pulse was 0.97. If we further assume that the
excitation probability is one, which is justifiable because we
performed the experiments at saturation, we can directly deduce the
collection efficiency of the setup from the measured count rate. We
reached a stable collection efficiency of $\eta=2.7$\,\% in
saturation, estimated from experimental collection efficiency at
10\,MHz laser rate, while having a total single photon count rate of
267\,kcts/s in saturation (see figure 4(b),(c)).

A more detailed analysis can be based on a two-level model. This
simple model describes the measured count rate $r(\Gamma)$ of the
laser repetition rate $\Gamma$. Assuming an excitation probability
of one which is justifiable as explained above, only the collection
efficiency $\eta$, the lifetime of the emitter $\gamma^{-1}$ and the
laser repetition rate $\Gamma$ influence the single photon count
rate

$$r(\Gamma) = \eta\cdot\Gamma\cdot\gamma\cdot\int_0^{1/\Gamma} e^{-\gamma t} \mathrm{d}t = \eta\cdot\Gamma\cdot(1-e^{-\gamma/\Gamma}).$$
We used this model for the fit in figure 4(c) where the count rate
in saturation is displayed as a function of the laser rate. From the
fit the photon collection efficiency $\eta$ can be deducted. A value
of $\eta=2.6$\,\% for the total collection efficiency is found in
agreement with  the estimated $\eta$ of 2.7\,\%. This is the highest
yet reported collection efficiency for a stable N-V defect centre in
diamond. Furthermore collection efficiency for the blinking defect
centre with count rates up to 2.4\,Mcts/s at CW excitation (see
figure \ref{fig:3}(b)) was even higher. It also emits ultra-high
single photon flux when excited with pulsed laser light. At laser
repetition rates of 10\,MHz and 385\,$\mu$W up to 420\,kcts/s were
detected. This determines a collection efficiency of 4.2\,\% on a
timescale of several ten seconds. The determined setup efficiency
$\eta_\mathrm{Setup}$ after the flat surface of the SIL of
$\eta_\mathrm{Setup}<0.23$ gives a source efficiency $\epsilon$,
i.e. fraction of light collected by the first objective, of
$\epsilon
>11.7$\,\% and $\epsilon >18.3$\,\% for the stable and the blinking
defect centre respectively. $\eta_\mathrm{Setup}$ was calculated
from reflection and transmission parameters of all optical elements
according to their data sheet values.

\begin{figure}
  \centering
  \includegraphics[width=1.0\textwidth]{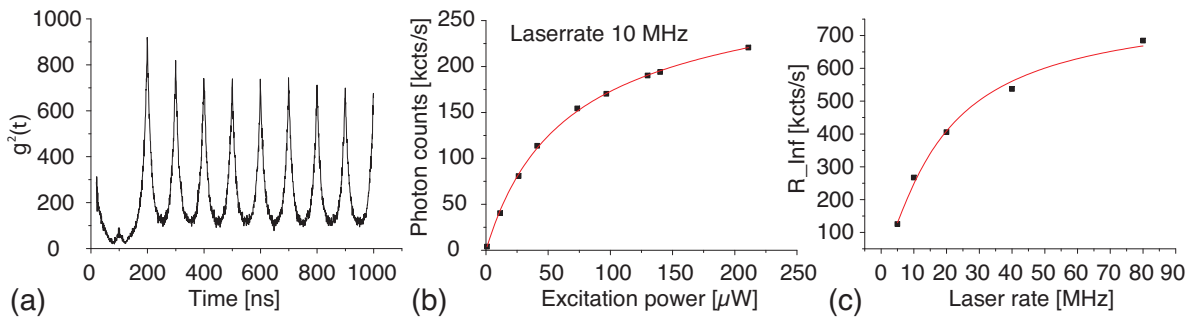}
  \caption{Pulsed autocorrelation function, saturation measurement at pulsed excitation, and emission in saturation as function of laser rate.
  (a) Pulsed autocorrelation measurement with $g^{(2)}(0)=0.16$
   of the stable, bright emitter depicted in figure \ref{fig:3(a)} excited at excitation intensity
  of 77\,$\mu$W with 10\,MHz laser rate and count rates of 221\,kcts/s. (b) Saturation curve of the same emitter as in (a)
  under pulsed excitation of 10\,MHz with R$_\mathrm{Inf} = 267\,$kcts/s and $I_\mathrm{Sat} = 60\,\mu$W. The red curve is a theoretical
  fit to the data (see text). (c) $R_\mathrm{Inf}$ as function of laser repetition rate. The theoretical fit to the data (red curve) is further
  explained in the text.}
  \label{fig:4}
\end{figure}

Single photon emission character of the emitters excited with pulsed
laser light was also proven measuring the normalized intensity
autocorrelation function $g^{(2)}(\tau)$. Figure \ref{fig:4}(a)
depicts the pulsed autocorrelation function of the N-V centre with
$g^{(2)}(0)=0.16$ at excitation intensity of 77\,$\mu$W and count
rates of 221\,kcts/s. Emission under pulsed excitation generally
provides a better suppression of background photons compared to CW
excitation as excitation radiation is limited to periods when the
electron is in the ground state, i.e. can be excited and emit a
single photon. The bright blinking defect centre had
$g^{(2)}(0)=0.11$ while excited with laser rates of 10\,MHz and
385\,$\mu$W excitation power.

\vspace{\baselineskip}

In conclusion, we report on an experimentally simple, but efficient
way to strongly enhance collection of single photons emitted by N-V
centres in diamond at room temperature. We integrated a ZrO$_2$
solid immersion lens with spin-coated nanodiamonds on its flat
surface into our confocal microscope. We measured count rates in
saturation of up to 853\,kcts/s for a stable N-V centre and up to
2.4\,Mcts/s for a blinking defect centre. Furthermore, our compact
SIL design provides access to about 100 N-V centres that emit more
than 400\,kcts/s. The overall collection efficiency of our setup is
up to 4.2\,\% while having a source efficiency $\epsilon$ of up to
$\epsilon >18.3$\,\%, opening the way towards much more efficient
diamond based on-demand single photon sources. Furthermore the setup
is so versatile that any kind of defect centre located in a
nanodiamond can be implemented thus allowing to integrate even
brighter emitters with smaller bandwidth. For a single Si-V centre
with a lifetime of 1.2\,ns as was just very recently presented
\cite{Neu2010} we would expect count rates up to 10\,Mcts/s. Also
cryogenic experiments will be possible as the immersion microscope
works oil free leading off towards two-photon interference
experiments.

\section*{Acknowledgement}
We thank T. Aichele for fruitful discussions. Financial support of
the BMBF (KEPHOSI) is acknowledged.


\section*{References}

\begin{thebibliography}{10}

\bibitem{O'Brien2009}
O'Brien J~L, Furusawa A and Vuckovic J 2009 Photonic quantum
technologies
  \textit{Nat. Photon.} 3 687--95

\bibitem{Neumann2010}
Neumann P, Beck J, Steiner M, Rempp F, Fedder H, Hemmer P~R,
Wrachtrup J and
  Jelezko F 2010 Single-Shot Readout of a Single Nuclear Spin \textit{Science}
  329 542--4

\bibitem{Dutt2007}
Dutt M~V~G, Childress L, Jiang L, Togan E, Maze J, Jelezko F, Zibrov
A~S,
  Hemmer P~R and Lukin M~D 2007 Quantum register based on individual electronic
  and euclear spin qubits in diamond \textit{Science} 316 1312--6

\bibitem{Bouwmeester1997}
Bouwmeester D, Pan J~W, Mattle K, Eibl M, Weinfurter H and Zeilinger
A 1997
  Experimental quantum teleportation \textit{Nature} 390 575--9

\bibitem{Schmid2009}
Schmid C, Kiesel N, Weber U~K, Ursin R, Zeilinger A and Weinfurter H
2009
  Quantum teleportation and entanglement swapping with linear optics logic
  gates \textit{New J. Phys.} 11 033008

\bibitem{PhysRevLett.89.067901}
Kuhn A, Hennrich M and Rempe G 2002 Deterministic single-photon
source for
  distributed quantum networking \textit{Phys. Rev. Lett.} 89 067901

\bibitem{Darquie2005}
Darquie B, Jones M~P~A, Dingjan J, Beugnon J, Bergamini S, Sortais
Y, Messin G,
  Browaeys A and Grangier P 2005 Controlled single-photon emission from a
  single trapped two-level atom \textit{Science} 309 454--6

\bibitem{Keller2004}
Keller M, Lange B, Hayasaka K, Lange W and Walther H 2004 Continuous
generation
  of single photons with controlled waveform in an ion-trap cavity system
  \textit{Nature} 431 1075--8

\bibitem{Lounis2000}
Lounis B and Moerner W~E 2000 Single photons on demand from a single
molecule
  at room temperature \textit{Nature} 407 491--3

\bibitem{Michler2000}
Michler P, Kiraz A, Becher C, Schoenfeld W~V, Petroff P~M, Zhang L,
Hu E and
  Imamoglu A 2000 A quantum dot single-photon turnstile device \textit{Science}
  290 2282--5

\bibitem{PhysRevLett.85.290}
Kurtsiefer C, Mayer S, Zarda P and Weinfurter H 2000 Stable
solid-state source
  of single photons \textit{Phys. Rev. Lett.} 85 290--3

\bibitem{springerlink:10.1140/epjd/e20020024}
Barnes W, Bj{\"o}rk G, Gérard J, Jonsson P, Wasey J, Worthing P and
Zwiller V
  2002 Solid-state single photon sources: light collection strategies
  \textit{EPJD} 18 197--210

\bibitem{Claudon2010}
Claudon J, Bleuse J, Malik N~S, Bazin M, Jaffrennou P, Gregersen N,
Sauvan C,
  Lalanne P and Gerard J~M 2010 A highly efficient single-photon source based
  on a quantum dot in a photonic nanowire \textit{Nat. Photon.} 4 174--7

\bibitem{Dousse2010}
Dousse A, Suffczynski J, Beveratos A, Krebs O, Lemaitre A, Sagnes I,
Bloch J,
  Voisin P and Senellart P 2010 Ultrabright source of entangled photon pairs
  \textit{Nature} 466 217--20

\bibitem{simpson:203107}
Simpson D~A, Ampem-Lassen E, Gibson B~C, Trpkovski S, Hossain F~M,
Huntington
  S~T, Greentree A~D, Hollenberg L~C~L and Prawer S 2009 A highly efficient two
  level diamond based single photon source \textit{Appl. Phys. Lett.} 94 203107

\bibitem{Wang2006}
Wang C, Kurtsiefer C, Weinfurter H and Burchard B 2006 Single photon
emission
  from SiV centres in diamond produced by ion implantation \textit{J. Phys. B}
  39 37

\bibitem{rabeau:134104}
Rabeau J~R, Huntington S~T, Greentree A~D and Prawer S 2005 Diamond
  chemical-vapor deposition on optical fibers for fluorescence waveguiding
  \textit{Appl. Phys. Lett.} 86 134104

\bibitem{wolters:141108}
Wolters J, Schell A~W, Kewes G, N{\"u}sse N, Schoengen M, Doscher H,
Hannappel
  T, Lochel B, Barth M and Benson O 2010 Enhancement of the zero phonon line
  emission from a single nitrogen vacancy center in a nanodiamond via coupling
  to a photonic crystal cavity \textit{Appl. Phys. Lett.} 97 141108

\bibitem{schroeder2010}
Schr{\"o}der T, Schell A~W, Kewes G, Aichele T and Benson O
Fiber-integrated
  diamond based single photon source \textit{Nano Lett.} under reviewing process

\bibitem{Beveratos2002}
Beveratos A, K{\"u}hn S, Brouri R, Gacoin T, Poizat J~P and Grangier
P 2002
  Room temperature stable single-photon source \textit{EPJD} 18 191--6

\bibitem{Babinec2010}
Babinec T~M, M H~J, Khan M, Zhang Y, Maze J~R, Hemmer P~R and Loncar
M 2010 A
  diamond nanowire single-photon source \textit{Nat. Nanotechnol.} 5 195--9

\bibitem{Hadden2010}
Hadden J~P, Harrison J~P, Stanley-Clarke A~C, Marseglia L, Ho Y~L~D,
Patton
  B~R, O'Brien J~L and Rarity J~G Strongly enhanced photon collection from
  diamond defect centres under micro-fabricated integrated solid immersion
  lenses \textit{arXiv:1006.2093v2}

\bibitem{Siyushev2010}
Siyushev P, Kaiser F, Jacques V, Gerhardt I, Bischof S, Fedder H,
Dodson J,
  Markham M, Twitchen D, Jelezko F and Wrachtrup J Integrated diamond optics
  for single photon detection \textit{arXiv:1009.0607v1}

\bibitem{Mansfield1990}
Mansfield S~M and Kino G~S 1990 Solid immersion microscope
\textit{Appl. Phys.
  Lett.} 57 2615--6

\bibitem{Terris1994}
Terris B~D, Mamin H~J, Rugar D, Studenmund W~R and Kino G~S 1994
Near-field
  optical data storage using a solid immersion lens \textit{Appl. Phys. Lett.}
  65 388--90

\bibitem{koyama:1667}
Koyama K, Yoshita M, Baba M, Suemoto T and Akiyama H 1999 High
collection
  efficiency in fluorescence microscopy with a solid immersion lens
  \textit{Appl. Phys. Lett.} 75 1667--9

\bibitem{Lukosz:77}
Lukosz W and Kunz R~E 1977 Light emission by magnetic and electric
dipoles
  close to a plane interface. I. Total radiated power \textit{J. Opt. Soc. Am.}
  67 1607--15

\bibitem{baba:6923}
Baba M, Sasaki T, Yoshita M and Akiyama H 1999 Aberrations and
allowances for
  errors in a hemisphere solid immersion lens for submicron-resolution
  photoluminescence microscopy \textit{J. Appl. Phys.} 85 6923--25

\bibitem{PhysRevB.79.235316}
Aharonovich I, Zhou C, Stacey A, Orwa J, Castelletto S, Simpson D,
Greentree
  A~D, Treussart F, Roch J~F and Prawer S 2009 Enhanced single-photon emission
  in the near infrared from a diamond color center \textit{Phys. Rev. B} 79
  235316

\bibitem{Steinmetz2010}
Steinmetz D, Neu E, Meijer J, Bolse W and Becher C Single photon
emitters based
  on Ni/Si related defects in single crystalline diamond
  \textit{arXiv:1007.0202v3}

\bibitem{Neu2010}
Neu E, Steinmetz D, Riedrich-Moeller J, Gsell S, Fischer M, Schreck
M and
  Becher C Single photon emission from silicon-vacancy centres in
  CVD-nano-diamonds on iridium \textit{arXiv:1008.4736v1}

\bibitem{Ladd2010}
Ladd T~D, Jelezko F, Laflamme R, Nakamura Y, Monroe C and O'Brien
J~L 2010
  Quantum computers \textit{Nature} 464 45--53


\end{thebibliography}


\end{document}